\documentclass[aps,showpacs,pra,superscriptaddress,]
{revtex4-2}

\usepackage{amsmath}
\usepackage{amssymb}
\usepackage{graphicx}
\usepackage{dcolumn}
\usepackage{bm}

\begin{document}

\title{Self-similar vector solitons for the coupled higher-order nonlinear
Schr\"{o}dinger equations in inhomogeneous optical fibers}
\author{ Houria Triki}
\affiliation{Radiation Physics Laboratory, Department of Physics, Faculty of Sciences,
	Badji Mokhtar University, P. O. Box 12, 23000 Annaba, Algeria}
\author{Vladimir I. Kruglov}
\affiliation{Centre for Engineering Quantum Systems, School of Mathematics and Physics, The University of Queensland, Brisbane, Queensland 4072, Australia}

\begin{abstract}
We prove the existence of two kinds of self-similar vector solitons in an
inhomogeneous optical fiber medium, where light propagation is governed by a
pair of coupled higher-order nonlinear Schr\"{o}dinger equations with
varying second- and third-order dispersions, self- and cross-phase
modulation nonlinearities, self-steepening, and linear gain/loss effects.
The newly found self-similar waves comprise bright-W-shaped and
kink-antikink waveforms with nonvanishing amplitudes. As a practical
example, we discuss the propagation dynamics of these soliton structures in
a periodically distributed fiber system as well as an exponential
dispersion-decreasing fiber. The results demonstrate that the parameter
functions of gain/loss and third-order dispersion serve as a key factor in
determining the nonlinear dynamics of self-similar vector solitons. In
particular, we find that precise control over the shape and dynamic
evolution of self-similar pulses can be achieved through a proper choice of
the distributed third-order dispersion parameter, while the gain/loss
coefficient controls their intensity.
\end{abstract}

\maketitle

\section{Introduction}

Soliton pulse propagation through Kerr-nonlinear media has been the focus of
continuous interest due to their great importance in ultrafast optical
communications and optical signal processing\ \cite{Agraw,Shao}. In optical
fibers, such nonlinear wave packets can be formed when group velocity
dispersion and self-phase modulation balance each other perfectly. While
dark solitons are produced in the normal dispersion regime \cite{Hasegawa2},
bright solitons are formed in the anomalous dispersion regime \cite%
{Hasegawa1}. A key feature of solitons is their capacity to preserve their
shape over long transmission distances. In addition to being observed in
optical fibers \cite{Mollenauer}, solitons have been demonstrated
experimentally in a variety of physical systems, such as photorefractive
crystals \cite{Photo1,Photo2}, femtosecond lasers \cite{Salin}, bulk optical
materials \cite{Aitchison}, and many others.

To describe the transmission of picosecond solitons, researchers generally
use the cubic nonlinear Schr\"{o}dinger equation, derived via the slowly
varying envelope approximation \cite{Agrawal}. Physically, this equation
describes how the envelope (or shape) of a pulse changes over distance as it
propagates through a nonlinear and dispersive medium, such as an optical
fiber. For the achievement of higher bit rates, it is advantageous to employ
ultrashort (femtosecond) pulses. Not only do these ultrashort pulses havc
great potential in several areas of application such as measurements of
ultrafast physical processes, infrared time-resolved spectroscopy and
sampling systems, but they are also invaluable for communications \cite%
{Agrawal,Alka}. However, results have demonstrated that as pulse durations
get shorter and peak powers increase, the effects of various physical
phenomena on short-pulse transmission and generation become significant \cite%
{Agraw}. Notably, if subpicosecond or femtosecond optical pulses are
injected, third-order dispersion and self-steepening processes become
important and they strongly influence the pulse dynamics \cite{Truta}. For
instance, the dispersion of the third order causes dispersive wave
radiation, pulse-shape distortion and spectral asymmetry \cite{Akhmediev}.
Additionally, the self-steepening effect causes the leading edge of an
optical pulse to become more sharp \cite{Anderson}. It is worth noting that
such process becomes significant for short-pulse transmission over long
distances \cite{Anderson,Tzoar}. Taking into account of these higher-order
processes, the evolution of a nonlinear short pulse envelope in an optical
fiber is governed by the higher-order nonlinear Schr\"{o}dinger (HNLS)
equation \cite{Zhao,Radha,Kru}:%
\begin{equation}
i\psi _{Z}-\frac{\beta _{2}}{2}\psi _{TT}+\gamma \left\vert \psi \right\vert
^{2}\psi -\frac{i\beta _{3}}{6}\psi _{TTT}+i\sigma (\left\vert \psi
\right\vert ^{2}\psi )_{T}=0,  \label{1}
\end{equation}

\noindent where $Z$ is the normalized propagation distance, $T$ is the
retarded time, $\psi $ is the complex envelope of the electric field, $\beta
_{2}$ and $\beta _{3}$ represent the coefficients of second- and third-order
dispersions,\ $\gamma $ is the fiber nonlinearity coefficient associated
with self-phase modulation, and $\sigma $ is the self-steepening coefficient.

In many practical applications, however, solitons often propagate
simultaneously in multiple fields with different frequencies or different
polarizations. Generally, these multi-wave dynamics are governed by coupled
nonlinear wave equations, which represent a natural extension of scalar
models \cite{H1,H2}. As an important extension, the HNLS equation (\ref{1})
is naturally generalized to the coupled form \cite{Radha}:%
\begin{equation}
i\psi _{1Z}-\frac{\beta _{2}}{2}\psi _{1TT}+\gamma \lbrack (\alpha
\left\vert \psi _{1}\right\vert ^{2}+\epsilon \left\vert \psi
_{2}\right\vert ^{2})\psi _{1}]-\frac{i\beta _{3}}{6}\psi _{1TTT}+i\sigma
\lbrack (\alpha \left\vert \psi _{1}\right\vert ^{2}+\epsilon \left\vert
\psi _{2}\right\vert ^{2})\psi _{1}]_{T}=0,  \label{2}
\end{equation}%
\begin{equation}
i\psi _{2Z}-\frac{\beta _{2}}{2}\psi _{2TT}+\gamma \lbrack (\epsilon
\left\vert \psi _{1}\right\vert ^{2}+\alpha \left\vert \psi _{2}\right\vert
^{2})\psi _{2}]-\frac{i\beta _{3}}{6}\psi _{2TTT}+i\sigma \lbrack (\epsilon
\left\vert \psi _{1}\right\vert ^{2}+\alpha \left\vert \psi _{2}\right\vert
^{2})\psi _{2}]_{T}=0,  \label{3}
\end{equation}%
where $\alpha \ $and $\epsilon $ are real constants.

Recently, bright and dark soliton solutions have been found for Eqs. (\ref{2}%
) and (\ref{3}) under the parametric conditions $\alpha =\epsilon $ and $%
\beta _{2}\sigma =$ $\gamma \beta _{3}$ in \cite{Radha}. In the case when $%
\beta _{3}=0$, these equations have been applied to examine the effects of
birefringence on light pulse propagation in the femtosecond range \cite{Hisakado}. Moreover, an inverse scattering formulation for Eqs. (\ref{2})
and (\ref{3}) with $\alpha =\epsilon $ and $\beta _{3}=0$ has been also
given in \cite{His}. In addition, breather solutions of the coupled HNLS
equations with $\alpha =\epsilon $ and $\beta _{3}=0$ have been recently
constructed through the traditional Darboux transformation in \cite{Ming}.
To the best of our knowledge, no existing studies have systematically
examined the existence and propagation dynamics of solitons in this model
when accounting for the inhomogeneities present in the nonlinear medium. In
this paper, we demonstrate for the first time the existence of two distinct
types of vector solitons in a more realistic fiber system with inhomogeneity.

The paper consists of the following. Section II introduces two novel kinds
of exact soliton pair solutions for the model equations with constant
coefficients. In Sec. III, we present a similarity transformation reducing
the inhomogeneous coupled HNLS equations with distributed coefficients and
gain or loss to the related constant-coefficients ones. Section IV is
devoted to the construction of self-similar wave solutions of the
generalized coupled HNLS equations with distributed coefficients governing
the light propagation in presence of the inhomogeneities of fiber system. In
Sec. V, we investigate the evolutional dynamics of the obtained self-similar
pulses in a specified soliton control system. The conclusions of this paper
will then be summarized in Section VI.

\section{Exact solutions to coupled HNLS equations}

A physically relevant problem is to investigate the existence of envelope
solitons in physical systems involving two or more interacting optical
fields. Such soliton waveforms, when they exist, contribute significantly to
a better understanding of various nonlinear phenomena arising in such media
modeled by a coupled form of NLS equations. It is very important to point
out that coupled NLS equations are widely used to model ultrashort optical
pulse propagation in diverse systems ranging from birefringent fibers \cite%
{CN1}, photovoltaic photorefractive crystals \cite{CN}, to plasmas \cite{CN2}%
, and Bose-Einstein condensates \cite{CN3}. Therefore, finding soliton
solutions to such model equations is of great importance. In this work, we
are considering the coupled HNLS equations:%
\begin{equation}
i\psi _{1Z}-\frac{\beta _{2}}{2}\psi _{1TT}+\gamma \lbrack (\alpha
\left\vert \psi _{1}\right\vert ^{2}+3\alpha \left\vert \psi _{2}\right\vert
^{2})\psi _{1}]-\frac{i\beta _{3}}{6}\psi _{1TTT}+i\sigma \lbrack (\alpha
\left\vert \psi _{1}\right\vert ^{2}+3\alpha \left\vert \psi _{2}\right\vert
^{2})\psi _{1}]_{T}=0,  \label{4}
\end{equation}%
\begin{equation}
i\psi _{2Z}-\frac{\beta _{2}}{2}\psi _{2TT}+\gamma \lbrack (3\alpha
\left\vert \psi _{1}\right\vert ^{2}+\alpha \left\vert \psi _{2}\right\vert
^{2})\psi _{2}]-\frac{i\beta _{3}}{6}\psi _{2TTT}+i\sigma \lbrack (3\alpha
\left\vert \psi _{1}\right\vert ^{2}+\alpha \left\vert \psi _{2}\right\vert
^{2})\psi _{2}]_{T}=0,  \label{5}
\end{equation}

\noindent which correspond to $\epsilon =3\alpha $ in Eqs. (\ref{2}) and (%
\ref{3}), a case that\ contains substantial dynamic property.

We may ask whether the coupled HNLS equations (\ref{4}) and (\ref{5})
possess closed form soliton solutions or not. In this section, we search for
exact traveling-wave solutions by use of the ansatz method. For the optical
communication, envelope solitons are of primary importance. Here we
demonstrate that two kinds of vector solitons are possible, thus
illustrating the rich dynamics of the system.

To start with, we represent the field components in the form \cite{Krug,Zhou}%
,%
\begin{equation}
\psi _{1}(Z,T)=U(\eta )\exp [i(\kappa Z-\delta T+\theta _{0})],  \label{6}
\end{equation}%
\begin{equation}
\psi _{2}(Z,T)=V(\eta )\exp [i(\kappa Z-\delta T+\theta _{0})],  \label{7}
\end{equation}%
where $U(\eta )$\ and $V(\eta )$\ are real functions, $\eta =T-qZ$ is the
traveling coordinate, $q=v^{-1}$\ is the inverse velocity, $\delta $\
represent the frequency shift, $\kappa $ is the wave number, and $\theta _{0}
$ is the initial phase constant.

Direct substitution of these forms of $\psi _{1}$ and $\psi _{2}$ in Eq. (%
\ref{4}) and separation of the real and imaginary parts leads to the
following differential equations: 
\begin{equation}
\frac{\beta _{3}}{6}\frac{d^{3}U}{d\eta ^{3}}+(q-\beta _{2}\delta -\frac{%
\beta _{3}}{2}\delta ^{2}-3\alpha \sigma U^{2}-3\alpha \sigma V^{2})\frac{dU%
}{d\eta }-6\alpha \sigma UV\frac{dV}{d\eta }=0,  \label{8}
\end{equation}%
\begin{equation}
\frac{1}{2}(\beta _{2}+\beta _{3}\delta )\frac{d^{2}U}{d\eta ^{2}}+(\kappa -%
\frac{\beta _{2}}{2}\delta ^{2}-\frac{\beta _{3}}{6}\delta ^{3})U-\alpha
(\gamma +\sigma \delta )U^{3}-3\alpha (\gamma +\sigma \delta )V^{2}U=0.
\label{9}
\end{equation}

Furthermore, the insertion of Eqs. (\ref{6}) and (\ref{7}) into Eq. (\ref{5}%
) and separation of the real and imaginary parts yield the coupled
equations: 
\begin{equation}
\frac{\beta _{3}}{6}\frac{d^{3}V}{d\eta ^{3}}+(q-\beta _{2}\delta -\frac{%
\beta _{3}}{2}\delta ^{2}-3\alpha \sigma V^{2}-3\alpha \sigma U^{2})\frac{dV%
}{d\eta }-6\alpha \sigma UV\frac{dU}{d\eta }=0,  \label{10}
\end{equation}%
\begin{equation}
\frac{1}{2}\left( \beta _{2}+\beta _{3}\delta \right) \frac{d^{2}V}{d\eta
^{2}}+(\kappa -\frac{\beta _{2}}{2}\delta ^{2}-\frac{\beta _{3}}{6}\delta
^{3})V-\alpha (\gamma +\sigma \delta )V^{3}-3\alpha (\gamma +\sigma \delta
)U^{2}V=0.  \label{11}
\end{equation}%
Then, Eqs. (\ref{8}) and (\ref{10}) can be integrated once and leads to: 
\begin{equation}
\frac{\beta _{3}}{6}\frac{d^{2}U}{d\eta ^{2}}+(q-\beta _{2}\delta -\frac{%
\beta _{3}}{2}\delta ^{2})U-\alpha \sigma U^{3}-3\alpha \sigma V^{2}U=C_{1},
\label{12}
\end{equation}%
\begin{equation}
\frac{\beta _{3}}{6}\frac{d^{2}V}{d\eta ^{2}}+(q-\beta _{2}\delta -\frac{%
\beta _{3}}{2}\delta ^{2})V-\alpha \sigma V^{3}-3\alpha \sigma U^{2}V=C_{2},
\label{13}
\end{equation}%
with $C_{1}$\ and $C_{2}$\ being the integration constants. As we are
interested in solitary wave solutions, it is necessary to set $C_{1}=C_{2}=0$%
\ and then Eqs. (\ref{9}) and (\ref{11}) are equivalent to Eqs. (\ref{12})
and (\ref{13}) respectively provided that, 
\begin{equation}
\frac{\kappa -\frac{1}{2}\beta _{2}\delta ^{2}-\frac{1}{6}\beta _{3}\delta
^{3}}{\beta _{2}+\beta _{3}\delta }=\frac{3}{\beta _{3}}(q-\beta _{2}\delta -%
\frac{\beta _{3}}{2}\delta ^{2}),  \label{14}
\end{equation}%
\begin{equation}
\frac{\gamma +\sigma \delta }{\beta _{2}+\beta _{3}\delta }=\frac{3\sigma }{%
\beta _{3}}.  \label{15}
\end{equation}%
The latter equations give the following expressions for the frequency shift
and wave number $\delta $\ and $\kappa $: 
\begin{equation}
\delta =\frac{\gamma }{2\sigma }-\frac{3\beta _{2}}{2\beta _{3}},  \label{16}
\end{equation}%
\begin{equation}
\kappa =\frac{\beta _{2}}{2}\delta ^{2}+\frac{\beta _{3}}{6}\delta ^{3}+%
\frac{3}{\beta _{3}}(\beta _{2}+\beta _{3}\delta )(q-q_{0}),  \label{17}
\end{equation}%
where the parameter $q_{0}=\beta _{2}\delta +\frac{1}{2}\beta _{3}\delta
^{2} $ has been introduced for brevity. Therefore, Eqs. (\ref{9}), (\ref{11}%
), (\ref{12}) and (\ref{13}) reduce to the coupled nonlinear ordinary
differential equations:%
\begin{equation}
\frac{d^{2}U}{d\eta ^{2}}+aU+b\left( \alpha U^{3}+3\alpha V^{2}U\right) =0,
\label{18}
\end{equation}%
\begin{equation}
\frac{d^{2}V}{d\eta ^{2}}+aV+b\left( \alpha V^{3}+3\alpha U^{2}V\right) =0,
\label{19}
\end{equation}%
Here the parameters $a$\ and $b$\ are defined by%
\begin{equation}
a=\frac{6\left( q-q_{0}\right) }{\beta _{3}},~~~~b=-\frac{6\sigma }{\beta
_{3}},  \label{20}
\end{equation}

Now we introduce new ans\"{a}tze for solving Eqs. (\ref{18}) and (\ref{19}),
which enable us to find novel classes of soliton pair solutions within the
framework of the coupled HNLS equations (\ref{4}) and (\ref{5}). The various
characteristics and formation conditions of the coupled solitons will be
also discussed in detail. Here, we identified two distinct types of soliton
pair solutions of Eqs. (\ref{4}) and (\ref{5}).

\begin{description}
\item[\textbf{Type 1.}] 
\end{description}

We have found a first type of exact analytical soliton solutions for Eqs. (%
\ref{18}) and (\ref{19}) of the form:%
\begin{equation}
U\left( \eta \right) =A+B\,\mathrm{sech}[w(\eta -\eta _{0})],  \label{21}
\end{equation}%
\begin{equation}
V\left( \eta \right) =A-B\,\mathrm{sech}[w(\eta -\eta _{0})],  \label{22}
\end{equation}%
where%
\begin{equation}
w^{2}=-a,\quad A^{2}=-\frac{a}{4b\alpha },\quad B^{2}=2A^{2}.  \label{23}
\end{equation}

\begin{figure}[h]
\includegraphics[width=1.3\textwidth]{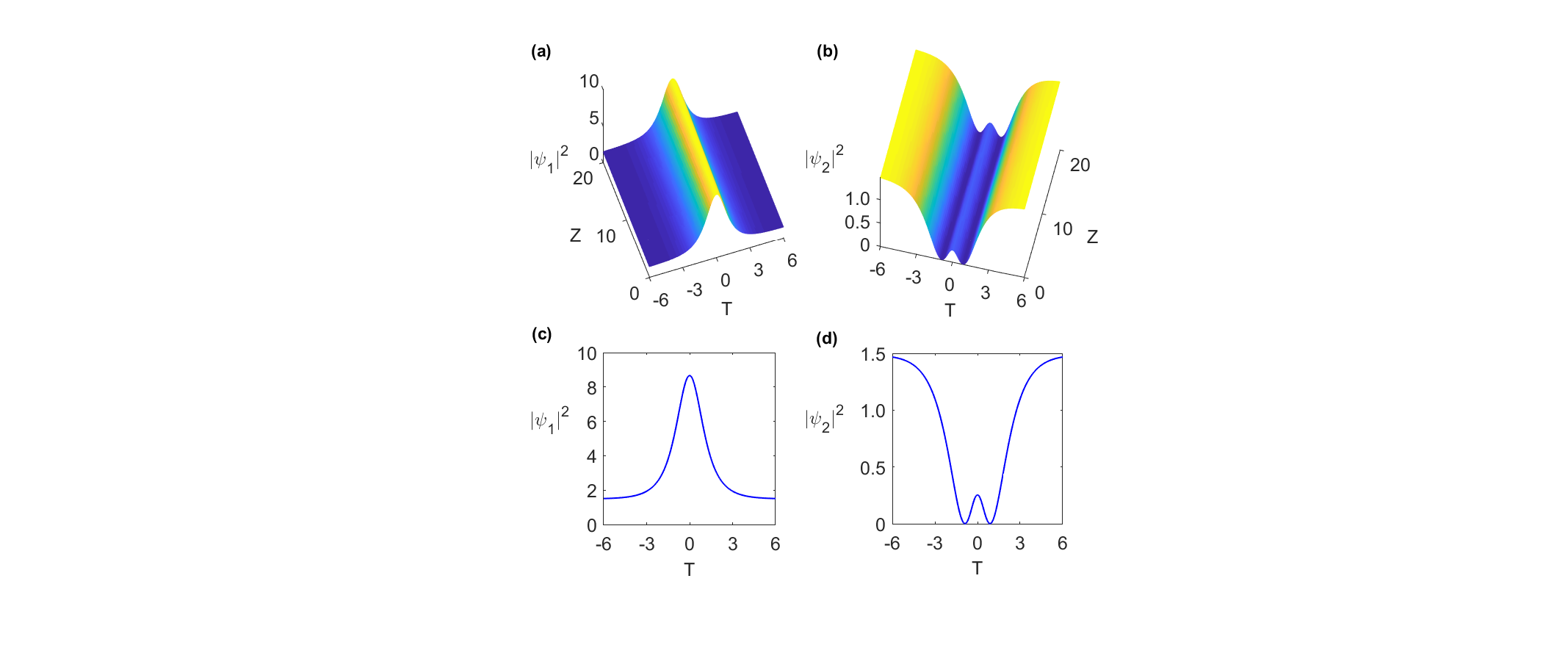}
\caption{Evolution of intensity wave profiles of the soliton
solutions (\ref{27}) and (\ref{28}) with parameters $\alpha =-0.02$,\ $\sigma =1$,\ $\beta _{2}=-0.471$, $\beta _{3}=0.6$,\ $\gamma =1.26$,\ $q=0.01$, $\eta_{0}=0$ (a) bright soliton solution (\ref{27}) and (b) W-shaped soliton solution (\ref{28}). Intensity profiles of (c) bright soliton solution (\ref{27}) and (d) W-shaped soliton solution (\ref{28}).}
\label{FIG.1.}
\end{figure}

\noindent Here the integration constant $\eta _{0}$ defines the position of
the localized pulse maximum. Then, utilizing Eqs. (\ref{20}) and (\ref{23}),
one finds that the pulse parameters $w$ and $A$ can be expressed as,%
\begin{equation}
w=\sqrt{-\frac{6\left( q-q_{0}\right) }{\beta _{3}}},  \label{24}
\end{equation}%
\begin{equation}
A=\pm \frac{1}{2}\sqrt{\frac{q-q_{0}}{\sigma \alpha }},  \label{25}
\end{equation}%
where $q$ is a free parameter. Note that we have for positive and negative
value of parameter $A$ two different solutions for the amplitude $B$ as,%
\begin{equation}
B=\sqrt{\frac{q-q_{0}}{2\sigma \alpha }},\quad B=-\sqrt{\frac{q-q_{0}}{%
2\sigma \alpha }},  \label{26}
\end{equation}%
because of relation $B^{2}=2A^{2}$.

Insertion of these results into Eqs. (\ref{6}) and (\ref{7}) and considering
the positive value of $B$, we obtain a novel class of exact solitary wave
solutions for the coupled HNLS equations (\ref{4}) and (\ref{5}) as%
\begin{equation}
\psi _{1}(Z,T)=A\left\{ 1+\sqrt{2}\,\mathrm{sech}\left[ w\left( T-qZ-\eta
_{0}\right) \right] \right\} e^{i(\kappa Z-\delta T+\theta _{0})},
\label{27}
\end{equation}%
\begin{equation}
\psi _{2}(Z,T)=A\left\{ 1-\sqrt{2}\,\mathrm{sech}\left[ w\left( T-qZ-\eta
_{0}\right) \right] \right\} e^{i(\kappa Z-\delta T+\theta _{0})},
\label{28}
\end{equation}

\noindent where $\delta $, $\kappa $, $w$, and $A$ are given by Eqs. (\ref%
{16}), (\ref{17}), (\ref{24}) and (\ref{25}), respectively.

We notice that the parameters of these soliton waveforms such as the wave
number $\kappa $, inverse width $w$, and amplitude $A$ depend on the fiber
parameters as well as a free parameter $q$. This enables us to achieve the
desired pulse power and width by carefully selecting the appropriate fiber
characteristics and free parameter $q$. One can also find from the relations
(\ref{24}), (\ref{25}) and (\ref{26}) that these localized waves exist when $%
\beta _{3}(q-q_{0})<0$ and $\sigma \alpha (q-q_{0})>0$.

Figure 1 depicts the intensity profiles of the soliton solutions (\ref{27})
and (\ref{28}) when $\alpha =-0.02$,\ $\sigma =1$,\ $\beta _{2}=-0.471$, $%
\beta _{3}=0.6$,\ $\gamma =1.26$,\ $q=0.01$, and $\eta _{0}=0$. Physically,
the solution (\ref{27}) represents a bright soliton envelope with a constant
background [see Figs. 1(a) and (c)] whereas the pulse represented by the
solution (\ref{28}) is a W-shaped soliton waveform [see Figs. 1(b) and (d)].
One can also see that the peak intensity of the W-shaped soliton waveform is
lower than that of the bright pulse. These results show that optical fibers
governed by the coupled HNLS equations (\ref{4}) and (\ref{5}) support the
existence of a W-shaped soliton structure. This finding is physically
significant because the occurrence of this kind of soliton in nonlinear
optics is relatively rare. We note that here the formation of this special
soliton structure arises from higher-order effects.

\begin{description}
\item[\textbf{Type 2.}] 
\end{description}

The second type of exact solutions we obtained for Eqs. (\ref{18}) and (\ref%
{19}) may be written as, 
\begin{equation}
U\left( \eta \right) =\Lambda +D\,\mathrm{tanh}[\mu (\eta -\eta _{0})],
\label{29}
\end{equation}%
\begin{equation}
V\left( \eta \right) =\Lambda -D\,\mathrm{tanh}[\mu (\eta -\eta _{0})],
\label{30}
\end{equation}%
where%
\begin{equation}
\mu ^{2}=\frac{a}{2},\quad \Lambda ^{2}=-\frac{a}{4b\alpha },\quad
D^{2}=\Lambda ^{2}.  \label{31}
\end{equation}

\begin{figure}[h]
\includegraphics[width=1.3\textwidth]{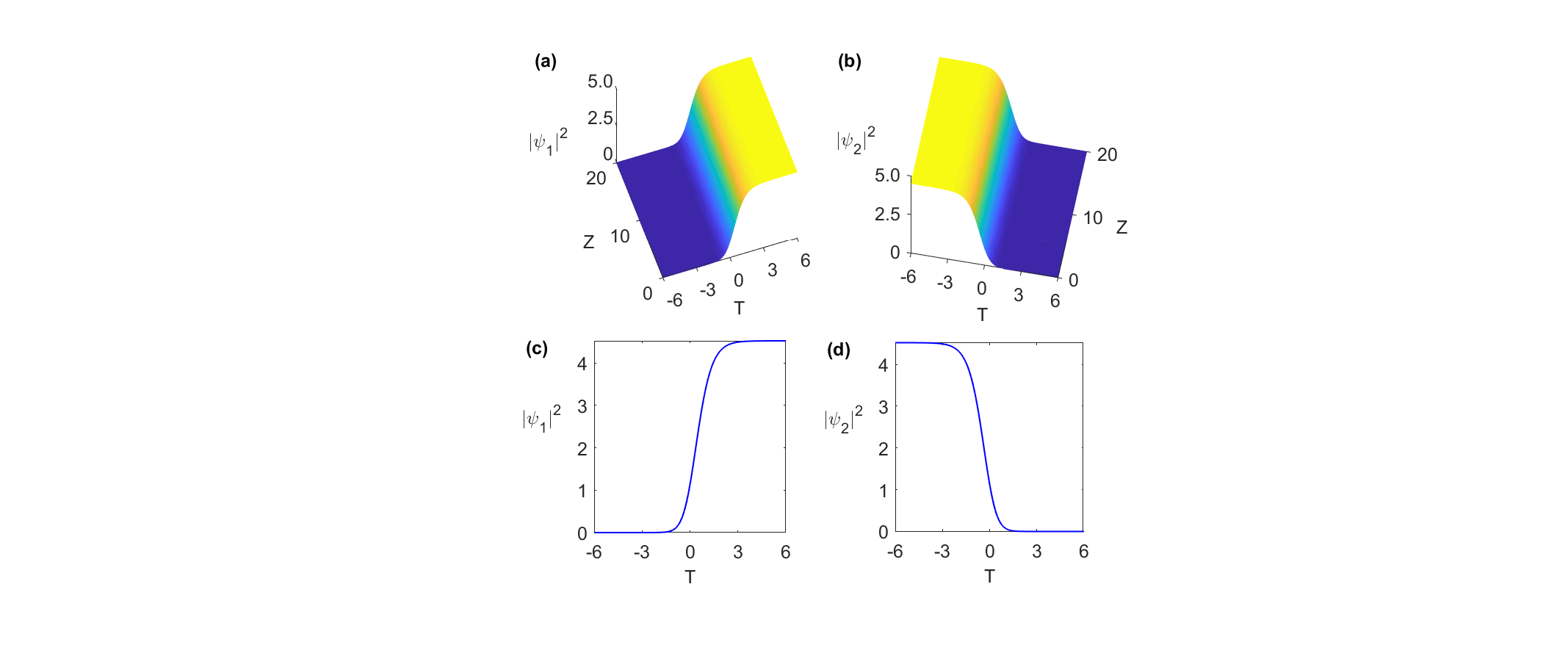}
\caption{Evolution of intensity wave profiles of the soliton
solutions (\ref{35}) and (\ref{36}) with parameters $\alpha =0.04$,\ $\sigma =1$,\ $\beta _{2}=-0.471$, $\beta _{3}=0.6$,\ $\gamma =-0.355$,\ $q=0.01$, $\eta_{0}=0$ (a) kink soliton solution (\ref{35}) and (b) antikink soliton solution (\ref{36}). Intensity profiles of (c) kink soliton solution (\ref{35}) and (d) antikink soliton solution (\ref{36}).}
\label{FIG.2.}
\end{figure}

Then, with use of Eqs. (\ref{20}) and (\ref{31}), one gets the following
expressions for the wave parameters $\mu $ and $\Lambda $: 
\begin{equation}
\mu =\sqrt{\frac{3\left( q-q_{0}\right) }{\beta _{3}}},  \label{32}
\end{equation}%
\begin{equation}
\Lambda =\pm \frac{1}{2}\sqrt{\frac{q-q_{0}}{\sigma \alpha }},  \label{33}
\end{equation}%
where $q$ is a free parameter. Note that we have for positive and negative
value of parameter $\Lambda $ two different solutions for the amplitude $D$
as 
\begin{equation}
D=\frac{1}{2}\sqrt{\frac{q-q_{0}}{\sigma \alpha }},\quad D=-\frac{1}{2}\sqrt{%
\frac{q-q_{0}}{\sigma \alpha }},  \label{34}
\end{equation}%
because of relation $D^{2}=\Lambda ^{2}$.

By substituting these results into Eqs. (\ref{6}) and (\ref{7}) and
considering the positive value of $D$, we find another class of exact
solitary wave solutions for the coupled HNLS equations (\ref{4}) and (\ref{5}%
) as:%
\begin{equation}
\psi _{1}(Z,T)=\Lambda \left\{ 1+\,\mathrm{tanh}\left[ \mu \left( T-qZ-\eta
_{0}\right) \right] \right\} e^{i(\kappa Z-\delta T+\theta _{0})},
\label{35}
\end{equation}%
\begin{equation}
\psi _{2}(Z,T)=\Lambda \left\{ 1-\,\mathrm{tanh}\left[ \mu \left( T-qZ-\eta
_{0}\right) \right] \right\} e^{i(\kappa Z-\delta T+\theta _{0})},
\label{36}
\end{equation}

\noindent which exist for material parameters obeying the parametric
conditions $\beta _{3}(q-q_{0})>0$ and $\sigma \alpha (q-q_{0})>0$.

Figure 2 shows examples of kink and antikink soliton solutions (\ref{35})
and (\ref{36}) for the following values of the parameters $\alpha =0.04$,\ $%
\sigma =1$,\ $\beta _{2}=-0.471$, $\beta _{3}=0.6$,\ $\gamma =-0.355$,\ $%
q=0.01$, and $\eta _{0}=0$. Note that these very physically relevant
nonlinear waves characteristically exist due to a balance among self- and
cross-phase modulation nonlinearities, second- and third-order dispersions,
and self-steepening effect.

Remarkably, the comparison between expressions (\ref{25}) and (\ref{33})
shows that the background amplitudes of the two types of vector solitons
determined above are identical. This implies that the background is
independent of the specific structure of the vector solitons.

\section{Similarity transformation of generalized coupled HNLS equations}

When optical pulses propagate through a waveguiding system, the
inhomogeneity of the medium affects the pulse transmission \cite{Meng},
while, the gain/loss influences the wave intensity \cite{Peacock}. A
challenging problem is the study of vector solitons with controllable
dynamics in inhomogeneous optical fibers within the femtosecond regime.
Achieving this control provides crucial insights into the properties of more
complex soliton pulse phenomena in optical fibers. If considering the
inhomogeneous nature of the fiber medium, the pulse propagation can be
described by the generalized coupled HNLS equations with varying
coefficients:%
\begin{equation}
i\Psi _{1z}-\frac{D_{2}(z)}{2}\Psi _{1tt}+R_{1}(z)[(\alpha \left\vert \Psi
_{1}\right\vert ^{2}+3\alpha \left\vert \Psi _{2}\right\vert ^{2})\Psi _{1}]-%
\frac{iD_{3}(z)}{6}\Psi _{1ttt}+iR_{2}(z)[(\alpha \left\vert \Psi
_{1}\right\vert ^{2}+3\alpha \left\vert \Psi _{2}\right\vert ^{2})\Psi
_{1}]_{t}=i\Gamma (z)\Psi _{1},  \label{37}
\end{equation}%
\begin{equation}
i\Psi _{2z}-\frac{D_{2}(z)}{2}\Psi _{2tt}+R_{1}(z)[(3\alpha \left\vert \Psi
_{1}\right\vert ^{2}+\alpha \left\vert \Psi _{2}\right\vert ^{2})\Psi _{2}]-%
\frac{iD_{3}(z)}{6}\Psi _{2ttt}+iR_{2}(z)[(3\alpha \left\vert \Psi
_{1}\right\vert ^{2}+\alpha \left\vert \Psi _{2}\right\vert ^{2})\Psi
_{2}]_{t}=i\Gamma (z)\Psi _{2},  \label{38}
\end{equation}%
where $D_{2}(z)$, $D_{3}(z)$, $R_{1}(z)$, $R_{2}(z)$ and $\Gamma (z)$ stand
for the distributed group velocity dispersion, third-order dispersion, cubic
nonlinearity, self-steepening, and linear gain/loss coefficients,
respectively.

In order to connect solutions of Eqs. (\ref{37}) and (\ref{38}) with those
of Eqs. (\ref{4}) and (\ref{5}), we apply the following transformations \cite%
{Dai1,Dai2}: 
\begin{equation}
\Psi _{1}(z,t)=\rho (z)e^{i\phi (z,t)}\psi _{1}\left[ Z(z),T(z,t)\right] ,
\label{39}
\end{equation}%
\begin{equation}
\Psi _{2}(z,t)=\rho (z)e^{i\phi (z,t)}\psi _{2}\left[ Z(z),T(z,t)\right] ,
\label{40}
\end{equation}%
where $\rho (z)$, $T(z,t)$, $Z(z)$ and $\phi (z,t)$ are real functions
representing the self-similar amplitude, self-similar time, effective
propagation distance, and self-similar phase variables of the waves,
respectively.
\begin{figure}[h]
\includegraphics[width=1.3\textwidth]{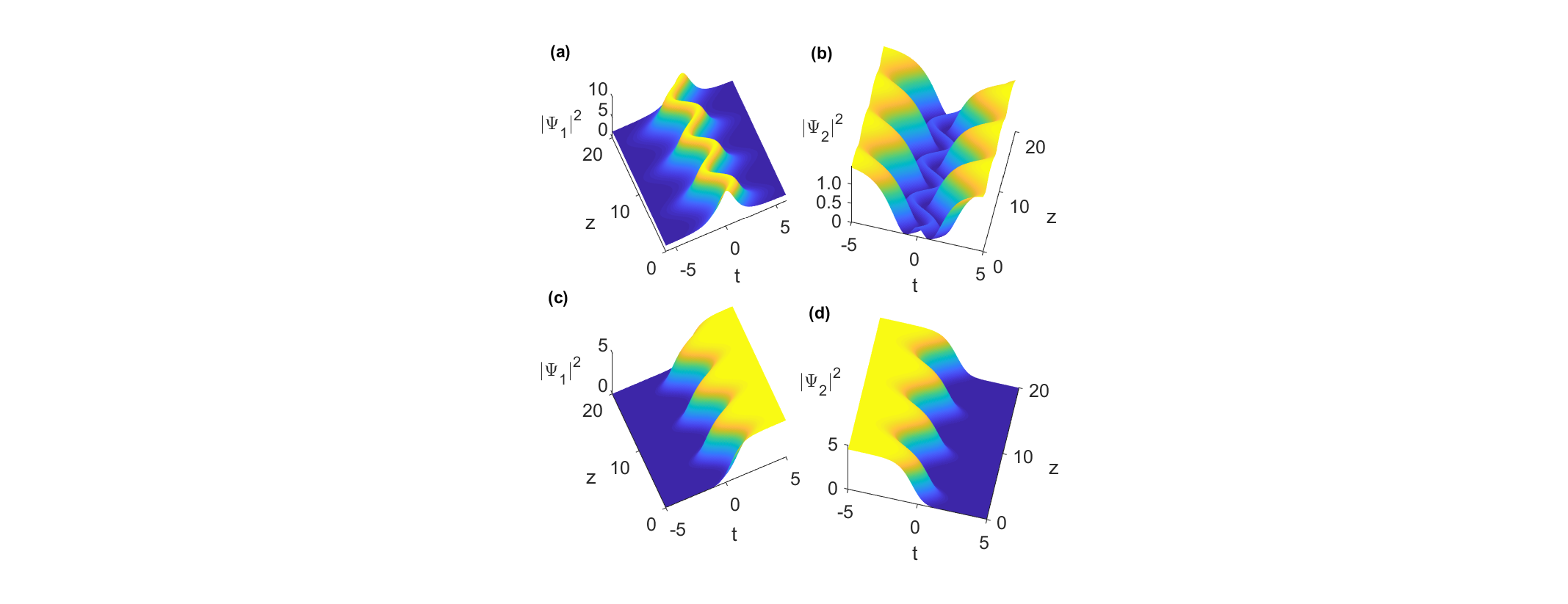}
\caption{Evolution of vector self-similar solitons with parameters $D_{3}(z)=D_{0}\cos (hz)$,\ $\Gamma (z)=\Gamma _{0}$, $D_{0}=3.31$,\ $h=1$, $k=1$, $\rho _{0}=1$,\ $p=1$, $\Gamma _{0}=0$, $\eta _{0}=t_{0}=0$ (a) self-similar bright wave (\ref{57}) (b) self-similar W-shaped wave (\ref{58}) (c) self-similar kink wave (\ref{61}) (d) self-similar antikink wave (\ref{62}). Other parameters are same as those in Figs. 1 and 2, respectively.}
\label{FIG.3.}
\end{figure}
\begin{figure}[h]
\includegraphics[width=1.3\textwidth]{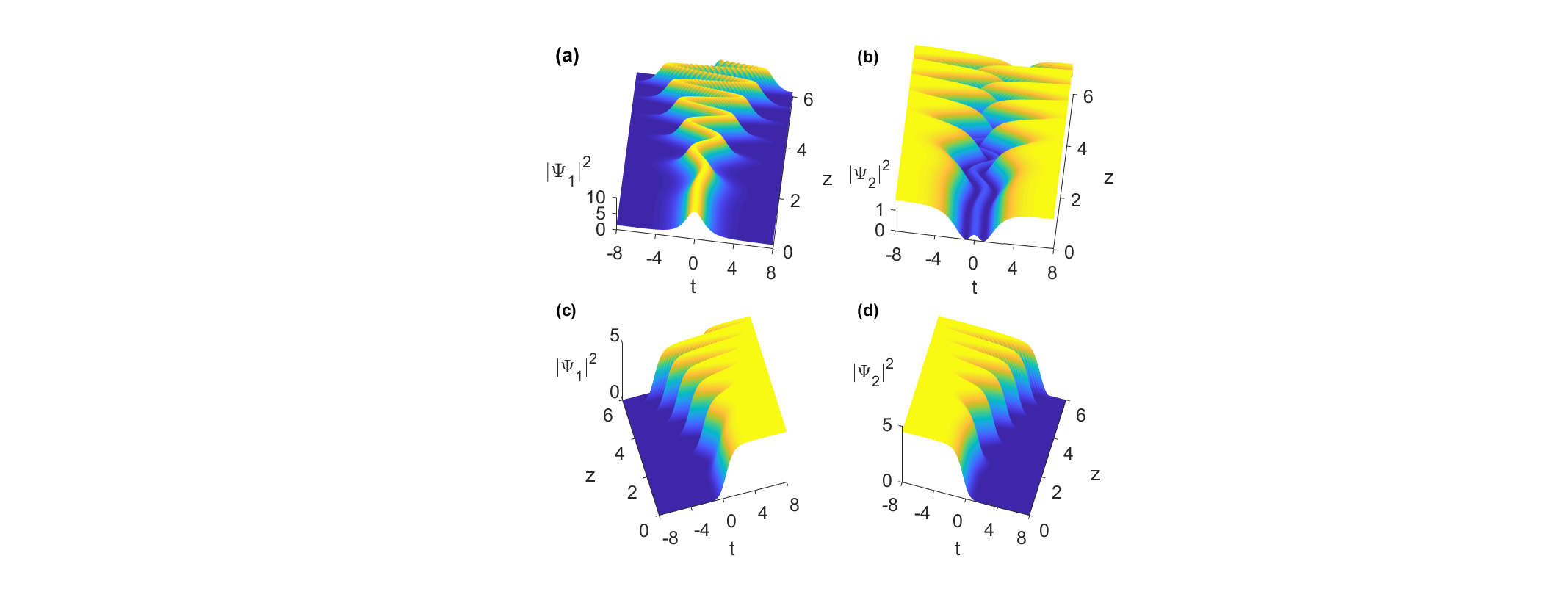}
\caption{Evolution of vector self-similar solitons with parameters $D_{3}(z)=D_{0}[z\cos (z^{2})-z^{3}\sin (z^{2})]$,\ $\Gamma (z)=\Gamma _{0}$, $D_{0}=-1$,\ $k=1$, $\rho _{0}=1$,\ $p=1$, $\Gamma _{0}=0$, $\eta_{0}=t_{0}=0$ (a) self-similar bright wave (\ref{57}) (b) self-similar W-shaped wave (\ref{58}) (c) self-similar kink wave (\ref{61}) (d) self-similar antikink wave (\ref{62}). Other parameters are same as those in Figs. 1 and 2, respectively. }
\label{FIG.4.}
\end{figure}
Substitution of Eqs. (\ref{39}) and (\ref{40}) into Eqs. (\ref{37}) and (\ref%
{38}) leads to Eqs. (\ref{4}) and (\ref{5}), but we must have the following
set of equations:%
\begin{equation}
\rho _{z}-\Gamma \rho -\frac{D_{2}}{2}\rho \phi _{tt}+\frac{D_{3}}{2}\rho
\phi _{t}\phi _{tt}=0,  \label{41}
\end{equation}%
\begin{equation}
T_{z}-D_{2}T_{t}\phi _{t}+\frac{D_{3}}{2}T_{t}\phi _{t}^{2}-\frac{D_{3}}{6}%
T_{ttt}=0,  \label{42}
\end{equation}%
\begin{equation}
\phi _{z}-\frac{D_{2}}{2}\phi _{t}^{2}-\frac{D_{3}}{6}\phi _{ttt}+\frac{D_{3}%
}{6}\phi _{t}^{3}=0,  \label{43}
\end{equation}%
\begin{equation}
\left( D_{3}\phi _{t}-D_{2}\right) T_{tt}+D_{3}T_{t}\phi _{tt}=0,\quad
T_{tt}=0,  \label{44}
\end{equation}%
\begin{equation}
\left( R_{1}-R_{2}\phi _{t}\right) \rho ^{2}=\gamma Z_{z},  \label{45}
\end{equation}%
\begin{equation}
\left( D_{2}-D_{3}\phi _{t}\right) T_{t}^{2}=\beta _{2}Z_{z},  \label{46}
\end{equation}%
\begin{equation}
D_{3}T_{t}^{3}=\beta _{3}Z_{z},\quad R_{2}T_{t}\rho ^{2}=\sigma Z_{z},
\label{47}
\end{equation}

We can solve this system of equations self-consistently and determine the
self-similar pulse parameters as:%
\begin{equation}
\rho (z)=\rho _{0}\exp \left( \int_{0}^{z}\Gamma (s)ds\right) ,  \label{48}
\end{equation}%
\begin{equation}
Z(z)=\frac{k^{3}}{\beta _{3}}\int_{0}^{z}D_{3}(s)ds,  \label{49}
\end{equation}%
\begin{equation}
T(z,t)=k\left[ t+\frac{p}{2}\left( \frac{2k\beta _{2}}{\beta _{3}}+p\right)
\int_{0}^{z}D_{3}(s)ds\right] +t_{0},  \label{50}
\end{equation}%
\begin{equation}
\phi (z,t)=p\left[ t+\frac{p}{6}\left( \frac{3k\beta _{2}}{\beta _{3}}%
+2p\right) \int_{0}^{z}D_{3}(s)ds\right] +\phi _{0},  \label{51}
\end{equation}%
along with the\ following constraints on the varying fiber parameters:%
\begin{equation}
R_{1}(z)=\frac{k^{2}\left( p\sigma +k\gamma \right) }{\beta _{3}\rho _{0}^{2}%
}D_{3}(z)\exp \left( -2\int_{0}^{z}\Gamma (s)ds\right) ,  \label{52}
\end{equation}%
\begin{equation}
R_{2}(z)=\frac{k^{2}\sigma }{\beta _{3}\rho _{0}^{2}}D_{3}(z)\exp \left(
-2\int_{0}^{z}\Gamma (s)ds\right) ,  \label{53}
\end{equation}%
\begin{equation}
D_{2}(z)=\left( p+\frac{k\beta _{2}}{\beta _{3}}\right) D_{3}(z),  \label{54}
\end{equation}%
where the real constants $k$ and $p$ are the parameters related to the width
and phase shift of the self-similar pulse, respectively, and the subscript $%
0 $ denotes the initial values of the corresponding parameters at distance $%
z=0 $.

Incorporating these results back into Eqs. (\ref{39}) and (\ref{40}), one
gets the general self-similar wave solutions to the inhomogeneous coupled
HNLS equations (\ref{37}) and (\ref{38}): 
\begin{eqnarray}
\Psi _{1}(z,t) &=&\rho _{0}\psi _{1}\left[ kt+\frac{1}{2}kp\left( \frac{%
2k\beta _{2}}{\beta _{3}}+p\right) \int_{0}^{z}D_{3}(s)ds+t_{0},\frac{k^{3}}{%
\beta _{3}}\int_{0}^{z}D_{3}(s)ds\right]  \notag \\
&&\times \exp \left( \int_{0}^{z}\Gamma (s)ds+i\phi (z,t)\right) ,~~~
\label{55}
\end{eqnarray}%
\begin{eqnarray}
\Psi _{2}(z,t) &=&\rho _{0}\psi _{2}\left[ kt+\frac{1}{2}kp\left( \frac{%
2k\beta _{2}}{\beta _{3}}+p\right) \int_{0}^{z}D_{3}(s)ds+t_{0},\frac{k^{3}}{%
\beta _{3}}\int_{0}^{z}D_{3}(s)ds\right]  \notag \\
&&\times \exp \left( \int_{0}^{z}\Gamma (s)ds+i\phi (z,t)\right) ,~~~
\label{56}
\end{eqnarray}%
where $\psi _{1}$ and $\psi _{2}$ are solutions of the constant-coefficient
coupled HNLS equations (\ref{4}) and (\ref{5}), while $\phi (z,t)$ is given
by the expression (\ref{51}).

A key step of constructing the exact analytical self-similar wave solutions
for the variable-coefficient coupled HNLS equations (\ref{37}) and (\ref{38}%
) via the transformations (\ref{55}) and (\ref{56}) is the derivation of
traveling waves of the constant-coefficient coupled HNLS equations (\ref{4})
and (\ref{5}).{\large \ }The above results show that Eqs. (\ref{4}) and (\ref%
{5}) admit two different types of exact solitary wave solutions. It is
worthy to{\large \ }note that Eqs. (\ref{52}), (\ref{53}) and (\ref{54}) are
the sufficient and necessary conditions for these self-similar solutions to
exist in the inhomogeneous Kerr optical medium.{\large \ }We also observe
from relations (\ref{49})-(\ref{51}) that the effective propagation
distance, similarity variable, and phase of the self-similar pulse are
fundamentally determined by the third-order dispersion parameter $D_{3}(z)$.
According to Eq. (\ref{48}), the self-similar amplitude $\rho (z)$ is
affected by the gain/loss distribution $\Gamma (z)$. Therefore, one can
manipulate the dynamics of self-similar soliton solutions by appropriately
choosing the two distributed parameters $D_{3}(z)$ and $\Gamma (z)$.

\section{Vector solitons for generalized coupled HNLS equation}

We now focus on the construction of exact self-similar vector soliton
solutions for the variable-coefficient coupled HNLS equations (\ref{37}) and
(\ref{38}), which describes the transmission of femtosecond pulses in the
optical fibers where all characteristic parameters vary with the propagation
distance $z$.

\begin{figure}[h]
\includegraphics[width=1.3\textwidth]{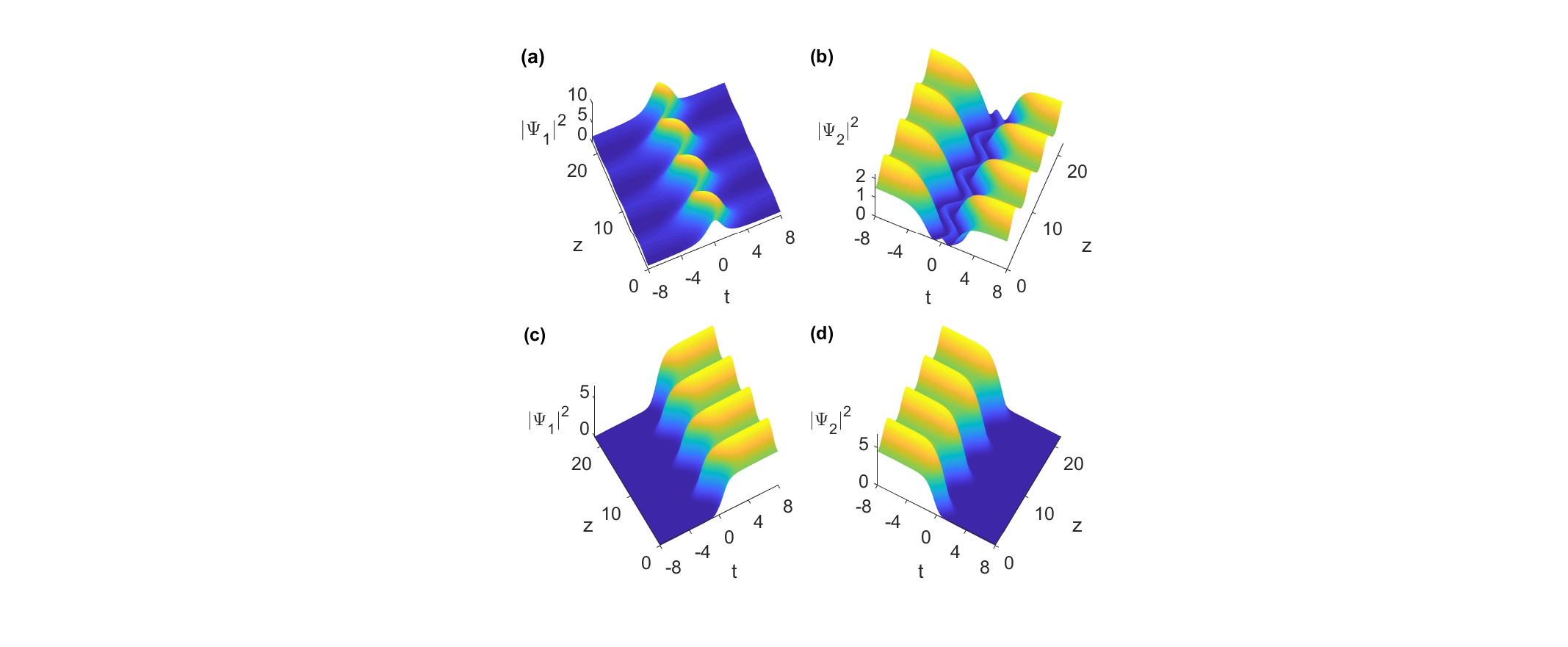}
\caption{Evolution of vector self-similar solitons with parameters $D_{3}(z)=D_{0}\cos (hz)$,\ $\Gamma (z)=\Gamma _{0}\mathrm{sin}(z)$, $D_{0}=3.31$,\ $h=1$, $\Gamma _{0}=0.1$, $k=1$, $\rho _{0}=1$,\ $p=1$, $\eta_{0}=t_{0}=0$ (a) self-similar bright wave (\ref{57}) (b) self-similar W-shaped wave (\ref{58}) (c) self-similar kink wave (\ref{61}) (d) self-similar antikink wave (\ref{62}). Other parameters are same as those in Figs. 1 and 2, respectively.}
\label{FIG.5.}
\end{figure}
\begin{figure}[h]
\includegraphics[width=1.3\textwidth]{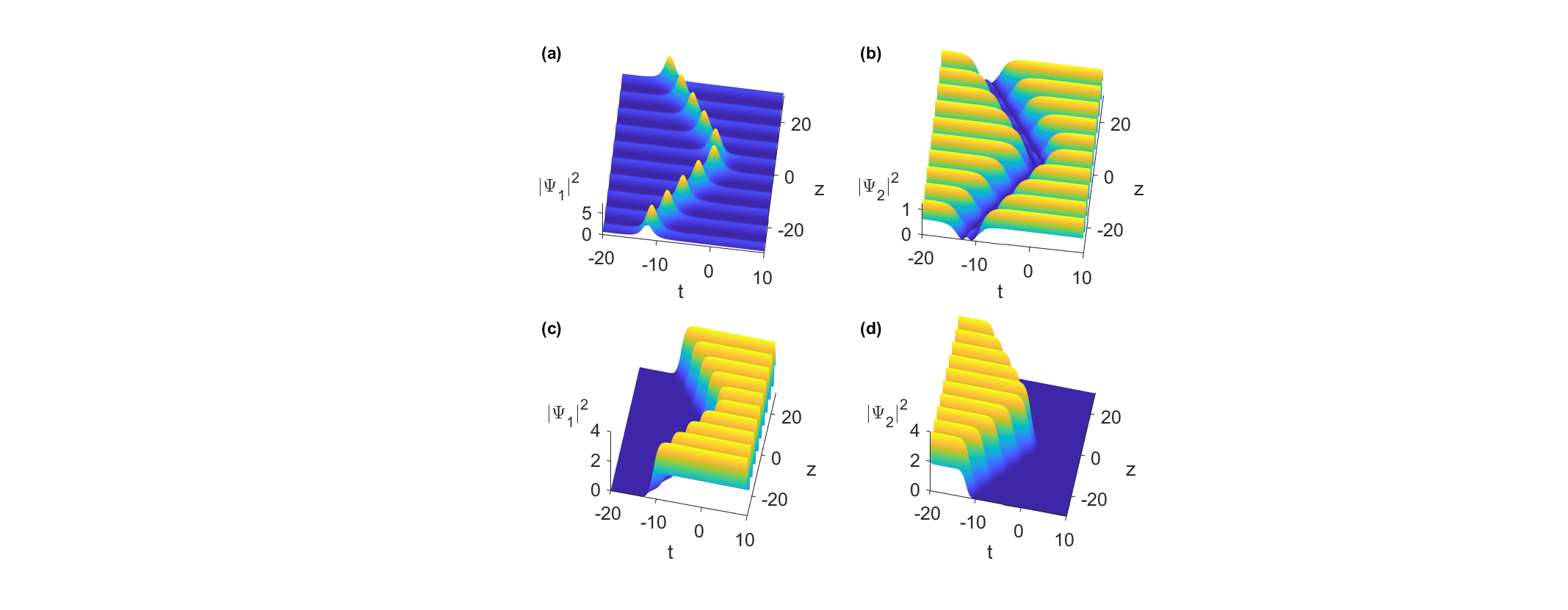}
\caption{Evolution of vector self-similar solitons with parameters $D_{3}(z)=D_{0}\mathrm{tanh}(z)$, $\Gamma (z)=\Gamma _{0}\sin (z)$, $D_{0}=-1.328$, $\Gamma_{0}=0.3$, $k=1$, $\rho _{0}=0.5$,\ $p=1$, $\eta_{0}=t_{0}=0$ (a) self-similar bright wave (\ref{57}) (b) self-similar W-shaped wave (\ref{58}) (c) self-similar kink wave (\ref{61}) (d) self-similar antikink wave (\ref{62}). Other parameters are same as those in Figs. 1 and 2, respectively.}
\label{FIG.6.}
\end{figure}
\begin{figure}[h]
\includegraphics[width=1.3\textwidth]{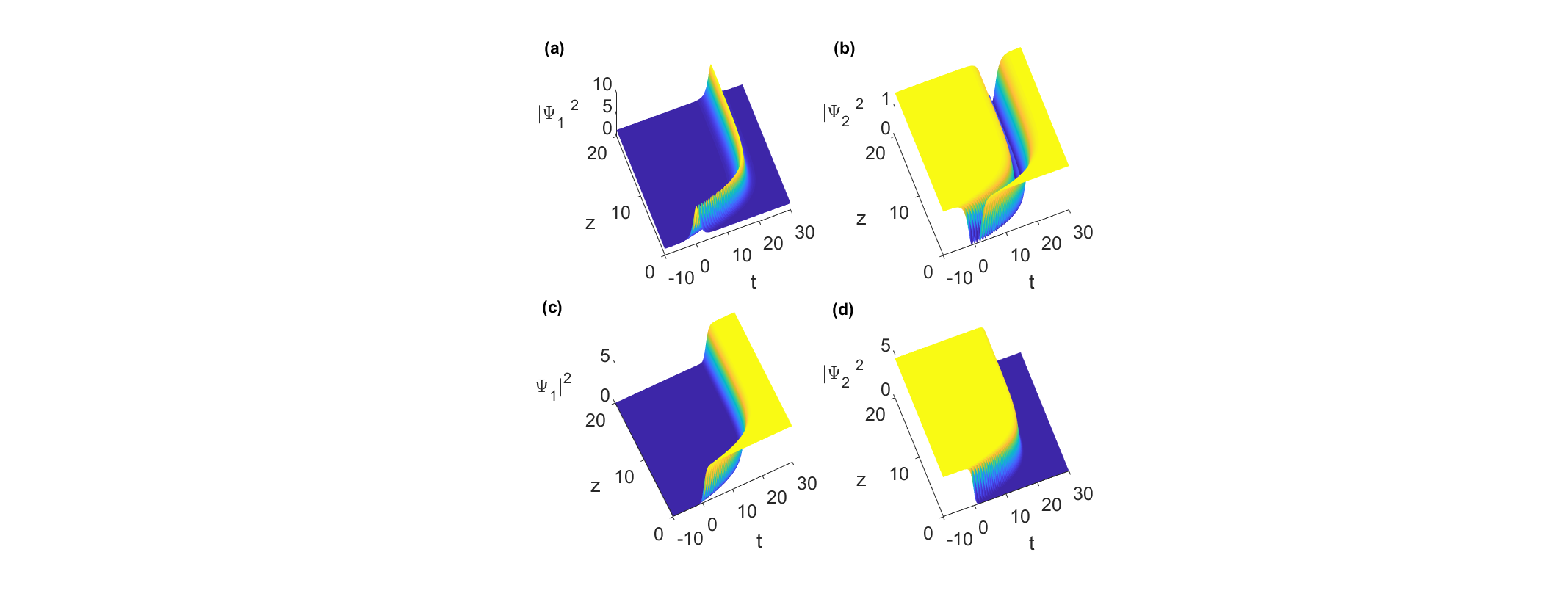}
\caption{Evolution of vector self-similar solitons with parameters $D_{3}(z)=D_{0}\exp (-g\,z)$, $\Gamma (z)=\Gamma _{0}$, $D_{0}=33.158$, $g=0.5$, $\Gamma _{0}=0$, $k=1$, $\rho _{0}=1$,\ $p=1$, $\eta _{0}=t_{0}=0$ (a) self-similar bright wave (\ref{57}) (b) self-similar W-shaped wave (\ref{58}) (c) self-similar kink wave (\ref{61}) (d) self-similar antikink wave (\ref{62}). Other parameters are same as those in Figs. 1 and 2, respectively.}
\label{FIG.7.}
\end{figure}

With use of the transformations (\ref{55}) and (\ref{56}) with (\ref{48})-(%
\ref{51}) and soliton solutions\ (\ref{27}) and (\ref{28}) of Eqs. (\ref{4})
and (\ref{5}), we obtain the first class of bright-W-shaped self-similar
wave solutions of Eqs. (\ref{37}) and (\ref{38}) in the form 
\begin{equation}
\Psi _{1}(z,t)=\rho _{0}A\exp \left( \int_{0}^{z}\Gamma (s)ds\right) \left\{
1+\sqrt{2}\,\mathrm{sech}\left( w\xi \right) \right\} \exp \left[ i\Theta
(z,t)\right] ,  \label{57}
\end{equation}%
\begin{equation}
\Psi _{2}(z,t)=\rho _{0}A\exp \left( \int_{0}^{z}\Gamma (s)ds\right) \left\{
1-\sqrt{2}\,\mathrm{sech}\left( w\xi \right) \right\} \exp \left[ i\Theta
(z,t)\right] ,  \label{58}
\end{equation}%
where the traveling wave variable $\xi $ is defined by%
\begin{equation}
\xi (z,t)=kt+\left[ \frac{kp}{2}\left( \frac{2k\beta _{2}}{\beta _{3}}%
+p\right) -\frac{qk^{3}}{\beta _{3}}\right] \int_{0}^{z}D_{3}(s)ds-\eta
_{0}+t_{0},  \label{59}
\end{equation}

\noindent and the total phase function $\Theta (z,t)$ reads,%
\begin{equation}
\Theta (z,t)=\kappa Z-\delta T+\phi (z,t)+\theta _{0},  \label{60}
\end{equation}

\noindent where $Z$ and $T$ are defined by Eqs. (\ref{49}) and (\ref{50})
respectively, the phase $\phi (z,t)$ is given by Eq. (\ref{51}), while $w$
and $A$ satisfy Eqs. (\ref{24}) and (\ref{25}).

A class of kink-antikink self-similar wave solutions of Eqs. (\ref{37}) and (\ref{38}) can be determined by substituting the soliton solutions\ (\ref{35}) and (\ref{36}) into the transformations (\ref{55}) and (\ref{56}) as
follows 
\begin{equation}
\Psi _{1}(z,t)=\rho _{0}\Lambda \exp \left( \int_{0}^{z}\Gamma (s)ds\right)
\left\{ 1+\,\mathrm{tanh}\left( \mu \xi \right) \right\} \exp \left[ i\Theta
(z,t)\right] ,  \label{61}
\end{equation}%
\begin{equation}
\Psi _{2}(z,t)=\rho _{0}\Lambda \exp \left( \int_{0}^{z}\Gamma (s)ds\right)
\left\{ 1-\mathrm{tanh}\left( \mu \xi \right) \right\} \exp \left[ i\Theta
(z,t)\right] ,  \label{62}
\end{equation}%
where $\xi $ and $\Theta (z,t)$ satisfy (\ref{59}) and (\ref{60}), while $%
\mu $ and $\Lambda $ are given by Eqs. (\ref{32}) and (\ref{33}). To the
best of our knowledge, the two self-similar soliton pairs presented above
for the variable-coefficient coupled HNLS equations (\ref{37}) and (\ref{38}%
) are reported here for the first time.

\section{The dynamics of bright-W-shaped and kink-antikink solitons}

We now examine the dynamics of the obtained self-similar waves in a
specified soliton control system. It is worthwhile to mention that for the
self-similar pulse solutions of Eqs. (\ref{37}) and (\ref{38}), only two of
the parameter functions $D_{2}(z)$, $R_{1}(z)$, $D_{3}(z)$, $R_{2}(z)$, and $%
\Gamma (z)$ in the inhomogeneous coupled HNLS equations (\ref{37}) and (\ref%
{38}) are free parameters. For example, if we choose $D_{3}(z)$ and $\Gamma
(z)$ as the free parameters, then the parameter functions $R_{1}(z)$, $%
R_{2}(z)$ and $D_{2}(z)$ will be obtained from Eqs. (\ref{52}), (\ref{53})
and (\ref{54}), respectively. To show controllable vector solitons, let us
begin with consideration of a soliton control system similarly to that in 
\cite{Serkin} i.e., the periodically distributed system \cite%
{Dai3,JZhang,RYang}:%
\begin{equation}
D_{3}(z)=D_{0}\cos (hz),\quad \Gamma (z)=\Gamma _{0},  \label{63}
\end{equation}%
where the parameters $D_{0}$\ and $h$\ are related to the third-order
dispersion and $\Gamma _{0}$\ stands for the constant net loss $\left(
<0\right) $ or gain $\left( >0\right) $. As concerns the\ other parameter
functions, they can be determined exactly using the constraint conditions (%
\ref{52}), (\ref{53}) and (\ref{54}). This choice is of practical importance
as it yields alternating regions of positive and negative dispersion and
nonlinearity, which is needed for an eventual stability of solitons \cite%
{Milo}. In Fig. 3, we depict the propagation dynamics of self-similar
bright-W-shaped and kink-antikink soliton solutions (\ref{57}), (\ref{58}), (\ref{61}) and (\ref{62}) with $D_{3}(z)$ and $\Gamma (z)$ given by Eq. (\ref{63}) for the parameters $D_{0}=3.31$,\ $h=1$, $k=1$, $\rho _{0}=1$,\ $p=1$, 
$\Gamma _{0}=0$, and $\eta _{0}=t_{0}=0$. One can see that the soliton pairs
exhibit a periodic oscillatory behavior during propagation through the
waveguide. For this periodic oscillatory evolution, referred to as
\textquotedblleft snakelike\textquotedblright\ \cite{Yan}, the profile of
the self-similar waves does not change as they propagates through the
optical medium, even though their position oscillates periodically. Unlike
the previous case, we can see that with the specific choice of the
dispersion function \cite{Trik} $D_{3}(z)=D_{0}[z\cos (z^{2})-z^{3}\sin
(z^{2})]$ with $D_{0}=-1$, the soliton structures are accelerated as they
propagate through the optical medium [see Fig. 4]. Hence, we can conclude
that in the absence of gain/loss management and under periodic dispersion,
self-similar solitons exhibit a snake-like propagation behaviour while
keeping their shape.

We now pose the important question: how can the gain/loss management
influence the evolutional dynamics of self-similar vector solitons in the
optical fiber medium? Let us consider this problem using the relevant
practical situation of a periodic gain/loss distribution which has recently
attracted a great deal of attention due to its promising applications in
optical communications \cite{Shally,TH}. To be precise, we consider a
periodically distributed fiber system with the dispersion and gain/loss
parameters \cite{JF}:%
\begin{equation}
D_{3}(z)=D_{0}\cos (hz),\quad \Gamma (z)=\Gamma _{0}\mathrm{sin}(z).
\label{64}
\end{equation}

Figure 5 displays the evolution behavior of self-similar soliton solutions (%
\ref{57}), (\ref{58}), (\ref{61}) and (\ref{62}) with $D_{3}(z)$ and $\Gamma
(z)$ given by Eq. (\ref{64}) for the parameter $D_{0}=3.31$,\ $h=1$, $\Gamma
_{0}=0.1$, $k=1$, $\rho _{0}=1$,\ $p=1$, and $\eta _{0}=t_{0}=0$. From this
figure, we observe that the intensity of the vector solitons oscillate
periodically compared to Figs. 3 and 4.

Next, we examine the evolution behavior of the bright W-shaped and
kink-antikink solitons [Eqs. (\ref{57}), (\ref{58}), (\ref{61}) and (\ref{62}%
)] in the soliton control system \cite{JF}:%
\begin{equation}
D_{3}(z)=D_{0}\mathrm{tanh}(z),\quad \Gamma (z)=\Gamma _{0}\mathrm{sin}(z).
\label{65}
\end{equation}

\noindent In this situation, the self-similar soliton pairs propagate along
a V-shaped trajectory, which is depicted in Fig. 6. From these results, we
may conclude that the distributed gain function $\Gamma (z)$ controls the
intensity of vector solitons, while dispersion $D_{3}(z)$ affects their
shape and propagation trajectory.

We now discuss the dynamics of the self-similar vector solitons in an
exponential dispersion decreasing fiber system with the distributed
dispersion parameter and constant gain/loss coefficient \cite{Dai3},%
\begin{equation}
D_{3}(z)=D_{0}\exp (-g\,z),\quad \Gamma (z)=\Gamma _{0},  \label{66}
\end{equation}

\noindent where $g$ and $D_{0}$ are the parameters to describe the group
velocity dispersion (with $g>0$ for dispersion decreasing fibers). 

Figure 7 displays the dynamical evolution of the vector soliton solutions (%
\ref{57}), (\ref{58}), (\ref{61}) and (\ref{62}) with $D_{3}(z)$ and $\Gamma
(z)$ given by Eq. (\ref{66}), taking the values of parameters as $%
D_{0}=33.158$, $g=0.5$, $\Gamma _{0}=0$, $k=1$, $\rho _{0}=1$,\ $p=1$, and $%
\eta _{0}=t_{0}=0$. We can see from the figure that this parametric choice
leads to the deceleration of self-similar vector solitons. 

It is worth mentioning that with other choices of gain/loss and third-order
dispersion profiles, we can obtain a wide variety of self-similar soliton
shapes, thus offering novel mechanisms for controlling ultrashort pulses in
complex optical systems.

\section{Conclusion}

To conclude, we have investigated the possible existence of vector soliton
pairs in an optical fiber possessing variations in dispersive and nonlinear
parameters along the propagation direction ($z$) of the waveguide. The
transmission process of ultrashort light pulses in such inhomogeneous fiber
media is described by a sytem of coupled higher-order nonlinear Schr\"{o}%
dinger equations with varying second- and third-order dispersions, self- and
cross-phase modulation nonlinearities, self-steepening, and linear gain/loss
effects. By applying the similarity transformation method, we reduced the
considered variable-coefficient sytem to the related constant-coefficient
one and derived exact analytical self-similar wave solutions. We have
identified two different kinds of exact analytical self-similar vector
solitons with nonvanishing amplitudes for the variable-coefficient sytem
which take the bright-W-shaped and kink-antikink structural profiles. As a
practical example, we have examined the dynamical behavior of these soliton
structures in a periodically distributed fiber system as well as an
exponential dispersion-decreasing fiber. The results showed that through a
suitable choice of the gain/loss and third-order dispersion profiles, we
could control the shape and propagation trajectory of the self-similar
vector solitons. These results offer profound theoretical insights into
nonlinear optical pulse dynamics in fiber media with higher-order effects
and practical guidelines for optimizing experimental pulse transmission
studies. Additionally, this study considerably advances the understanding of
transmission properties of optical waves in the presence of inhomogeneity
naturally existing in all real systems and lays a basis for controllable
wave management in optical fiber systems involving two interacting optical
fields.

\end{document}